\documentclass[a4paper,10pt,DIV=12,flalign, abstracton]{scrartcl}

\setkomafont{disposition}{\fontfamily{ptm}}%
\setkomafont{title}{\fontfamily{ptm}}%

\usepackage[utf8]{inputenc}
\usepackage[T1]{fontenc}
\usepackage{amssymb,amsmath,amsthm}
\usepackage{graphicx}
\usepackage{hyperref}
\usepackage{csquotes}
\usepackage{color}
\usepackage{subfig}

 \usepackage[sort&compress,numbers]{natbib}

\newcommand{\ndim}{n}
\newcommand{\surface}[1]{\partial{#1}}
\renewcommand{\d}[1]{\mathrm{d}{#1}}

\newcommand{\dsurf}{\;\d S}

\newcommand{\Kron}[1]{\delta_{#1}}

\newcommand{\matr}[1]{\left[#1\right]}
\newcommand{\partderivf}[3][{}]{\frac{\partial^{#1}{#2}}{\partial{#3}^{#1}}}
\newcommand{\matprod}{\cdot}

\newcommand{\inv}[1]{{#1}^{-1}}
\newcommand{\ttpermutcomp}{\epsilon}
\newcommand{\placeholder}{(\circ)}
\newcommand{\Lagrangemult}[1][{}]{\lambda^{#1}}
\newcommand{\Lagrangefunc}{\mathcal{L}}

\newcommand{\rate}[1]{\dot{#1}}
\newcommand{\stresscomp}{\sigma}
\newcommand{\displcomp}{u}
\newcommand{\straincomp}{\varepsilon}

\newcommand{\strainratecomp}{\rate{\straincomp}}
\newcommand{\Airyfunc}{F}

\newcommand{\strainenergy}[1][{}]{\mathcal{W}_{#1}}

\newcommand{\emod}{Y}
\newcommand{\Poissrat}{\nu}

\newcommand{\Lamesec}{\mu} %
\newcommand{\Lamefirst}{\lambda} %

\newcommand{\stressgradcomp}{R}

\newcommand{\microdisplcomp}{\Phi}

\newcommand{\microdispldevcomp}{\microdisplcomp}
\newcommand{\microdispldevratecomp}{\rate{\microdisplcomp}}

\newcommand{\gendisplcomp}{\Psi}
\newcommand{\gendisplratecomp}{\rate{\gendisplcomp}}

\newcommand{\strainmacroSGcomp}{E}
\newcommand{\strainmacroSGratecomp}{\rate{\strainmacroSGcomp}}
\newcommand{\stiffmatSG}{\hat{\matr{A}}}

\newcommand{\complmatSG}{\hat{\matr{B}}}
\newcommand{\complmatSGcomp}{\hat{B}}
\newcommand{\complcoeffSG}{\tilde{b}}
\newcommand{\stiffcoeffSG}{\tilde{a}}
\newcommand{\fluctuation}[1]{\Delta{#1}}
\newcommand{\displmacrocomp}{U}

\newcommand{\strainmacrocomp}{E}

\newcommand{\locmacrocomp}{X}

\newcommand{\locmicrocomp}{x}

\newcommand{\locmicrorelcomp}{y}

\newcommand{\stressmacrocomp}{\Sigma}
\newcommand{\strainenergymacro}[1][{}]{\overline{\strainenergy}_{#1}}
\newcommand{\strainenergycompmacro}[1][{}]{\overline{\strainenergy}_{#1}^{*}}

\newcommand{\powerintmacro}{P^{\mathrm{int}}}
\newcommand{\workfluxmacrocomp}{Q^{\mathrm{mech}}}
\newcommand{\averop}[2][{}]{\left\langle#2\right\rangle_{#1}} %

\newcommand{\domaincell}[1][{}]{\Delta V#1} %
\newcommand{\domaincellbound}[1][(\locmacro)]{\surface{\domaincell[]}#1}
\newcommand{\domaincellboundhalf}[1][(\locmacro)]{\surface{\domaincell[]}^+#1}

\newcommand{\ointcell}{\oint\limits_{\domaincellbound[]}\!\!}
\newcommand{\geommomcomp}{G}

\newcommand{\radcell}{r_{\mathrm{a}}}

\newcommand{\radvoid}{r_{\mathrm{i}}}
\newcommand{\VVF}{c}

\newcommand{\shearmodeff}{\mu^{\mathrm{(eff)}}}

\newcommand{\Lamefirsteff}{\lambda^{\mathrm{(eff)}}}
\newcommand{\emodeff}{Y^{\mathrm{(eff)}}}
\newcommand{\straintenstest}{\bar{\varepsilon}}
\newcommand{\tenstestpecwidth}{H}
\newcommand{\emodapparaent}{\emod_{\mathrm{app}}}
\newcommand{\charlengthSGtens}{\ell}

\newcommand{\inputsvg}[1]{\includegraphics{#1}}
\newcommand{\inputgnuplot}[1]{\includegraphics{#1}} %

\title{Kinematics and constitutive relations in the stress-gradient theory: interpretation by homogenization}

\author{Geralf Hütter, Karam Sab, Samuel Forest}

\begin{document}

\maketitle

\begin{abstract}
The stress-gradient theory has a third order tensor as kinematic degree of freedom, which is work-conjugate to the stress gradient. This tensor was called micro-displacements just for dimensional reasons. Consequently, this theory requires a constitutive relation between stress gradient and micro-displacements, in addition to the conventional stress-strain relation. The formulation of such a constitutive relation and identification of the parameters therein is difficult without an interpretation of the micro-displacement tensor.

The present contribution presents an homogenization concept from a Cauchy continuum at the micro-scale towards a stress-gradient continuum at the macro-scale. Conventional static boundary conditions at the volume element are interpreted as a Taylor series whose next term involves the stress gradient. A generalized Hill-Mandel lemma shows that the micro-displacements can be identified with the deviatoric part of the first moment of the microscopic strain field. Kinematic and periodic boundary conditions are provided as alternative to the static ones. The homogenization approach is used to compute the stress-gradient properties of an elastic porous material.
The predicted negative size effect under uni-axial loading is compared with respective experimental results for foams and direct numerical simulations from literature.

\end{abstract}

\section{Introduction}
\label{sec:introduction}

The classical Cauchy theory of continuum mechanics requires a constitutive relation between stress and strain. The constitutive parameters appearing therein can, for dimensional reasons, consequently always be grouped into those with dimension of a stress and dimensionless ones. Lacking an intrinsic length scale, this theory predicts a power-law scaling behavior when considering self-similar specimens of different size, the integer scaling exponent just depending on whether stress, strains, forces or displacements are considered. Deviations from this scaling behavior are termed \emph{size effects} and have been observed for numerous physical phenomena, cf.~\citep{Aifantis2003}.
That is why certain generalized theories of continuum mechanics have been proposed in the literature. A classification of the generalizations was given by \citet{Maugin2011}. Most of the generalized theories fall into the class of micro-morphic continua, which were established by \citet{Mindlin1964} and Eringen \citep{Eringen1964}. Therein, the (dimensionless) micro-deformation is introduced as additional kinematic degree of freedom. 
Certain sub-classes of theories, like the micro-polar theory (Cosserat theory) or the strain-gradient theory can be obtained by imposing kinematic constraints to the micro-deformation.
As alternative approach, \citet{Forest2012} imposed a kinetic constraint to obtain a stress-gradient theory. Therein, a kinematic degree of freedom $\microdisplcomp_{ijk}$ appears as work-conjugate quantity to the gradient 
\begin{align}
	\stressgradcomp_{ijk}:=\frac{\partial\stressmacrocomp_{ij}}{\partial\locmacrocomp_k}\,.    \label{eq:defstressgrad}
\end{align}
of the stress tensor $\stressmacrocomp_{ij}$.
Though, the stress gradient cannot take arbitrary values but it is restricted by the equilibrium conditions as will be detailed below. 
Due the presence of $\stressgradcomp_{ijk}$ in the respective potentials, Neumann boundary conditions do not involve only the tractions as normal component of $\stressmacrocomp_{ij}$, but \emph{all} components of $\stressmacrocomp_{ij}$ need to be prescribed at a surface. Alternative Dirichlet boundary conditions involve additional terms as well, cf.~\citep{Sab2016}.
The third-rank tensor $\microdisplcomp_{ijk}$ has the dimension of length, which is why it was termed \enquote{micro-displacements}.
Like all generalized theories of continuum mechanics, the stress-gradient requires additional constitutive relations. Their formulation and the interpretation of the boundary conditions is difficult without an interpretation of the tensor of micro-displacements $\microdisplcomp_{ijk}$.

The scope of the present contribution is to provide a homogenization methodology from a classical, but heterogeneous, continuum at the micro-scale towards a homogeneous stress-gradient theory at the macro-scale.

The present contribution is structured as follows: Section~\ref{sec:homogenization} presents the homogenization theory, before this theory is employed in
Section~\ref{sec:homogenization_foam} to compute the macroscopic non-classical constitutive parameters of a plane elastic micro-structure with pores. These constitutive parameters are used in Section~\ref{sec:uniaxial_tension} to predict the size effect under uni-axial tension. Finally, Section~\ref{sec:summary} closes with a summary and conclusions.

\section{Homogenization theory}
\label{sec:homogenization}

\begin{figure}
	\centering
	\subfloat[]{\inputsvg{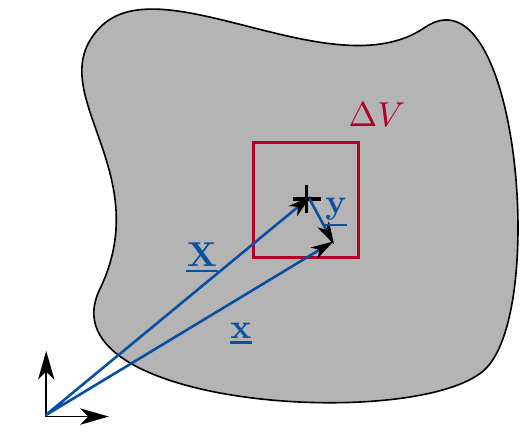} \label{fig:volumeelement}} \hspace{1cm}
	\subfloat[]{\inputsvg{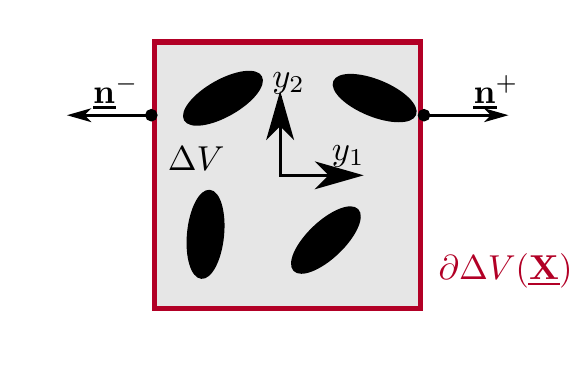} \label{fig:periodic_BCs} }
	\caption[Homogenization procedure]{Homogenization procedure: \subref{fig:volumeelement} volume element, \subref{fig:periodic_BCs} heterogeneous microstructure \citep{Huetter2017,Huetter2018Cosserat}}
	\label{fig:homogenization}
\end{figure}

In a homogenization procedure, a material with heterogeneous micro-structure is replaced by an homogeneous continuum with (more or less) equivalent macroscopic properties. For this purpose, a volume element $\domaincell$ is considered, which contains the relevant heterogeneities of the micro-structure as sketched in \figurename~\ref{fig:homogenization}.
In the following, capital symbols refer to macroscopic quantities, and lower-case symbols to microscopic ones. For instance, $\stresscomp_{ij}$ and $\straincomp_{ij}$ are the microscopic stress and strain, respectively, whereas $\stressmacrocomp_{ij}$ and $\strainmacrocomp_{ij}$ refer to their macroscopic counterparts.

In the classical theory of homogenization by \citet{Hill1963}, either kinematic boundary conditions $\displcomp_i=\strainmacrocomp_{ij}\locmicrorelcomp_j$ can be prescribed for the displacements on $\domaincellbound$, or static ones $n_i\stresscomp_{ij}=n_i\stressmacrocomp_{ij}$ for the tractions. Therein, $\locmicrorelcomp_j=\locmicrocomp_j-\locmacrocomp_j$ refers to the position vector of a point $\locmicrocomp_j$ relative to the center $\locmacrocomp_j=\averop{\locmicrocomp_j}$ of the volume element, cmp.\ \figurename~\ref{fig:homogenization}. The operator $\averop{\placeholder}$ computes the volume average over the volume element $\domaincell[]$.

\begin{figure}
	\centering
	\subfloat[]{\inputsvg{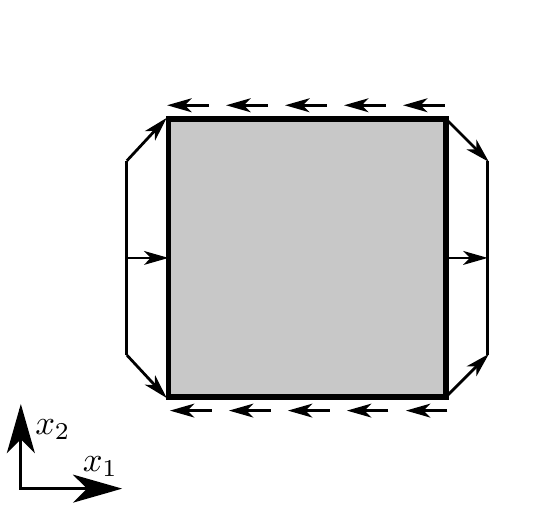} \label{fig:cell_Rload_R111}}
	\subfloat[]{\inputsvg{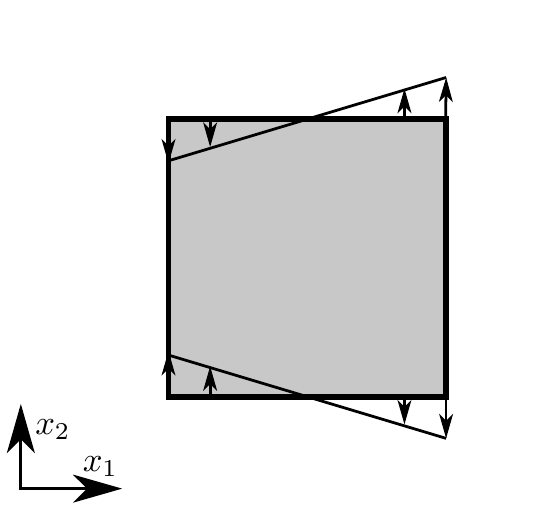} \label{fig:cell_Rload_R221}}
	\caption[Loading to the volume element by macroscopic stress gradients]{Loading to the volume element by macroscopic stress gradients: \subref{fig:cell_Rload_R111} $\stressgradcomp_{111}=-\stressgradcomp_{122}=-\stressgradcomp_{212}$, \subref{fig:cell_Rload_R221} $\stressgradcomp_{221}$ }
	\label{fig:cell_Rload}
\end{figure}

Gologanu, Kouznetsova et al.\ \citep{Gologanu1997,Kouznetsova2002} interpreted the kinematic boundary conditions as a Taylor series. In this sense, they incorporated an additional term to Hill's expression to obtain a homogenization scheme for the strain-gradient theory.
\citet{Muehlich2012} argued that an analogous expansion of Hill's static boundary condition would yield the homogenization for a stress-gradient theory. This proposal shall be exploited here in detail. Using the notation of the stress-gradient theory, an expanded \emph{static boundary condition} thus reads
\begin{equation}
  \stresscomp_{ij}n_i=n_i\left[\stressmacrocomp_{ij}+\stressgradcomp_{ijk}\locmicrorelcomp_k\right]\quad\forall\locmicrorelcomp_k\in\domaincellbound\,.
	\label{eq:staticBCstressgrad}
\end{equation}
Purely static boundary conditions are prone to the condition that, in absence of volume forces $\stresscomp_{ij,i}=0$, prescribed tractions need to be self-equilibrating (statically admissible): 
\begin{align}
  \ointcell \stresscomp_{ij}n_i\dsurf&=0, &
	\ointcell \stresscomp_{ij}n_i\locmicrorelcomp_k \ttpermutcomp_{jkl}\dsurf&=0\,.
\end{align}
For the particular tractions~\eqref{eq:staticBCstressgrad} these conditions require
\begin{align}
	\stressgradcomp_{iji}&=\stressmacrocomp_{ij,i}=0,& \stressmacrocomp_{ij}&=\stressmacrocomp_{ji}\,, \label{eq:equilibriummacro}
\end{align}
corresponding to the macroscopic equilibrium conditions.\footnote{This approach is used in many textbooks and lectures to derive the equilibrium conditions~\eqref{eq:equilibriummacro} for the Cauchy theory.}
Consequently, the stress gradient is symmetric $\stressgradcomp_{ijk}=\stressgradcomp_{jik}$ and deviatoric in the sense $\stressgradcomp_{ijj}=\stressgradcomp_{jij}=0$.
The loading to a volume element by stress gradients according to Eq.~\eqref{eq:staticBCstressgrad} is shown schematically in \figurename{~\ref{fig:cell_Rload}}.

Furthermore, a homogenization theory requires a condition of macro-homogeneity (Hill-Mandel condition), which defines the macroscopic mechanical power $\powerintmacro$ as average over its microscopic pendant:  
\begin{equation}
	\averop{\stresscomp_{ij}\strainratecomp_{ij}}=\powerintmacro(\locmacrocomp_k)\,.
	\label{eq:HillMandel}
\end{equation}
By partial integration, the left-hand side of Eq.~\eqref{eq:HillMandel} can be transformed to a surface integral over the boundary $\domaincellbound$ of the volume element $\domaincell$ into which boundary condition~\eqref{eq:staticBCstressgrad} can be inserted. After rearrangement and application of the divergence theorem, the left-hand side of Eq.~\eqref{eq:HillMandel} becomes
\begin{equation}
  \averop{\stresscomp_{ij}\strainratecomp_{ij}}=\frac{1}{\domaincell}\ointcell\stresscomp_{ij}n_i\rate{\displcomp}_j \dsurf =\stressmacrocomp_{ij}\averop{\strainratecomp_{ij}} +\stressgradcomp_{ijk}\left[\averop{\strainratecomp_{ij}\locmicrorelcomp_k}-\frac{1}{\ndim+1}\left(\averop{\strainratecomp_{im}\locmicrorelcomp_m}\Kron{jk}+\averop{\strainratecomp_{jm}\locmicrorelcomp_m}\Kron{ik}\right)\right]
	\label{eq:HillMandelinserted}
\end{equation}
Therein, $\ndim=\Kron{kk}$ refers to the dimension of space ($\ndim=2$ or $\ndim=3$).
From Eq.~\eqref{eq:HillMandelinserted}, a strain tensor $\strainmacroSGcomp_{ij}$ and a third-order tensor $\microdispldevcomp_{ijk}$, called tensor of micro-displacements \citep{Forest2012}, can be introduced as work-conjugate macroscopic deformation measures to $\stressmacrocomp_{ij}$ and $\stressgradcomp_{ijk}$, respectively, as
\begin{align}
  \strainmacroSGcomp_{ij}&=\averop{\straincomp_{ij}}=\frac{1}{2\domaincell}\ointcell \displcomp_i n_j+\displcomp_j n_i \dsurf \label{eq:micromacrostrain} \\
  \microdispldevcomp_{ijk}
	=&\averop{\straincomp_{ij}\locmicrorelcomp_k}-\frac{1}{\ndim+1}\left(\averop{\straincomp_{im}\locmicrorelcomp_m}\Kron{jk}+\averop{\straincomp_{jm}\locmicrorelcomp_m}\Kron{ik}\right) \label{eq:micromacromicrodispl} \\
	=&\frac{1}{2\domaincell}\ointcell\left(\displcomp_i n_j+\displcomp_j n_i\right)\locmicrorelcomp_k-\frac{1}{\ndim+1}\left(\displcomp_i n_m+\displcomp_m n_i\right)\locmicrorelcomp_m\Kron{jk}-\frac{1}{\ndim+1}\left(\displcomp_j n_m+\displcomp_m n_j\right)\locmicrorelcomp_m\Kron{ik}\dsurf\,. \nonumber
\end{align}
Thereby, it was taken into account that the stress and stress gradient exhibit symmetries, and so do their work-conjugate quantities ($\strainmacroSGcomp_{ij}=\strainmacroSGcomp_{ji}$, $\microdispldevcomp_{ijk}=\microdispldevcomp_{jik}$ and $\microdispldevcomp_{ijj}=0$).
Equation~\eqref{eq:micromacromicrodispl} indicates that the micro-displacement tensor $\microdispldevcomp_{ijk}$ corresponds to the deviatoric part of the first moment of the local strain field.
Furthermore, it shall be mentioned that micro-macro relations~\eqref{eq:micromacrostrain} and \eqref{eq:micromacromicrodispl} are objective, i.~e., that they are invariant to superimposed rigid-body motions. 

For a hyperelastic material $\stresscomp_{ij}\strainratecomp_{ij}=\dot{\strainenergy}$, Eq.~\eqref{eq:HillMandelinserted} can be integrated in time to a macroscopic strain energy potential
\begin{equation}
   \strainenergymacro(\strainmacroSGcomp_{ij},\microdispldevcomp_{ijk})=\averop{\strainenergy(\straincomp_{ij})}
	 \label{eq:strainenergymacro}
\end{equation}
with
\begin{align}
	\stressmacrocomp_{ij}&=\partderivf{\strainenergymacro}{\strainmacroSGcomp_{ij}}, &
	\stressgradcomp_{ijk}&=\partderivf{\strainenergymacro}{\microdispldevcomp_{ijk}}\,.
\end{align}

Furthermore, it is required that the energy at the macroscopic scale is conserved. This means, that it must be possible to convert the internal power $\powerintmacro=\stressmacrocomp_{ij}\strainmacroSGratecomp_{ij}+\stressgradcomp_{ijk}\microdispldevratecomp_{ijk}$ to the divergence of a flux $\workfluxmacrocomp_{i}$ of mechanical power:
\begin{equation}
	\powerintmacro=\workfluxmacrocomp_{i,i}\,.
\end{equation}
By partial integration of $\powerintmacro$ using the equilibrium conditions~\eqref{eq:equilibriummacro} and the definition~\eqref{eq:defstressgrad} of the stress gradient, it turns out that the flux of mechanical work has to be identified as
\begin{equation}
  \workfluxmacrocomp_{i}=\stressmacrocomp_{ij}\rate{\displmacrocomp}_j+\stressmacrocomp_{jk}\microdispldevratecomp_{jki}\,.
\label{eq:workfluxmacro}
\end{equation}
Therein, $\displmacrocomp_j(\locmacrocomp_k)$ is the macroscopic displacement field. Furthermore, the kinematic relation for the strain becomes
\begin{equation}
	\strainmacroSGcomp_{ij}=\displmacrocomp_{(i,j)}+\microdispldevcomp_{ijk,k}\,.
	\label{eq:strainmacro}
\end{equation}
The round brackets $_{(ij)}$ are used here and in the following to indicate the symmetric part of a tensor with respect to indices $i$ and $j$.
The kinematic relation~\eqref{eq:strainmacro} involves the divergence of the micro-displacements at the right-hand side, in addition to the symmetric part of the displacement gradient. 

In this context, it may be recalled, that it is an (implicit) ad-hoc postulate of the classical homogenization theory of \citet{Hill1963}, that the strain field $\strainmacroSGcomp_{ij}$ is macroscopically compatible, i.~e., that it it is related to a macroscopic displacement field via a kinematic relation, and that the field of macroscopic stresses satisfies equilibrium conditions. In the present approach, both, the macroscopic equilibrium conditions~\eqref{eq:equilibriummacro} as well as the kinematic relation~\eqref{eq:strainmacro} are an outcome of the homogenization procedure.

Alternatively, relations~\eqref{eq:workfluxmacro} and \eqref{eq:strainmacro} can be written in terms of a \enquote{generalized displacement tensor} \citep{Sab2016}   
\begin{equation}
	\gendisplcomp_{ijk}:=\frac{1}{2}\left(\displmacrocomp_i\Kron{jk}+\displmacrocomp_j\Kron{ik}\right)+\microdisplcomp_{ijk}
	\label{eq:gendisplacements}
\end{equation}
in short as $\strainmacroSGcomp_{ij}=\gendisplcomp_{ijk,k}$ and $\workfluxmacrocomp_{i}=\stressmacrocomp_{jk}\gendisplratecomp_{jki}$, respectively. The trace $\gendisplcomp_{ijj}=(\ndim+1)\displmacrocomp_i/2$ is directly related to the macroscopic displacement vector, whereas the deviatoric part of $\gendisplcomp_{ijk}$ corresponds to the micro-displacement tensor $\microdisplcomp_{ijk}$.
In view of Eq.~\eqref{eq:micromacromicrodispl}$_2$, the micro-macro relation for the generalized displacement tensor is formulated as
\begin{align}
	\gendisplcomp_{ijk}=&\frac{1}{2\domaincell}\ointcell\left(\displcomp_i n_j+\displcomp_j n_i\right)\locmicrorelcomp_k \dsurf
	=\averop{\straincomp_{ij}\locmicrorelcomp_k}+\frac12\left(\averop{\displcomp_i}\Kron{jk}+\averop{\displcomp_j}\Kron{ik}\right)
	\label{eq:micromacrogendispl}\,.
\end{align}
The deviatoric part of Eq.~\eqref{eq:micromacrogendispl} is identical to Eq.~\eqref{eq:micromacromicrodispl} for the micro-displacement. Furthermore, for a superimposed rigid translation the right-hand side of \eqref{eq:micromacrogendispl} transforms according to Eq.~\eqref{eq:gendisplacements}. 

\begin{figure}
	\centering
	\subfloat[]{\inputsvg{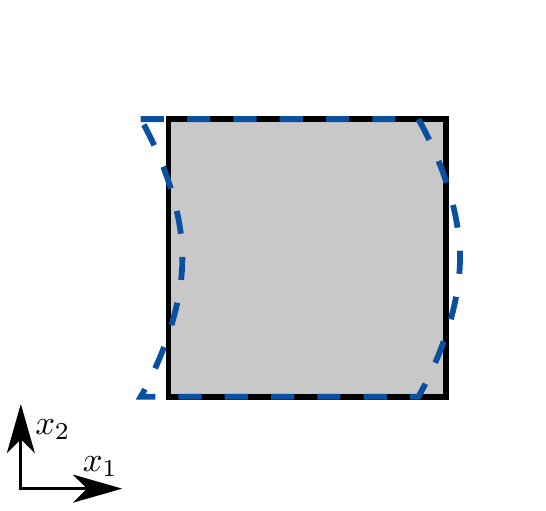} \label{fig:cell_Phi111}}
	\subfloat[]{\inputsvg{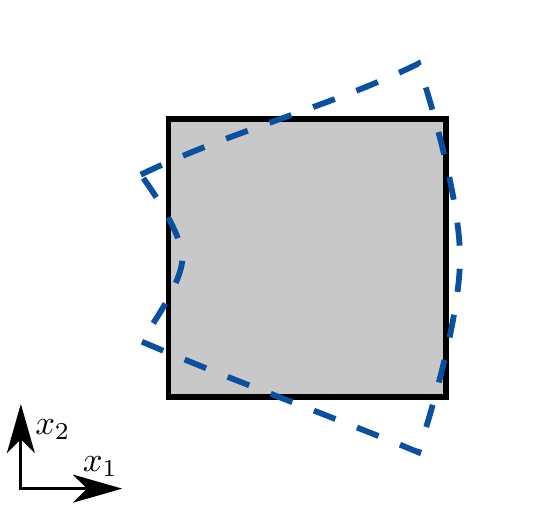} \label{fig:cell_Phi221}}
	\caption[]{Non-classical kinematic boundary conditions: \subref{fig:cell_Phi111} $\microdispldevcomp_{111}=-\microdispldevcomp_{122}=-\microdispldevcomp_{212}$, \subref{fig:cell_Phi221} $\microdispldevcomp_{221}$ }
	\label{fig:cell_microdispl_kinematicBC}
\end{figure}

In classical homogenization, kinematic or periodic boundary are usually favored over static ones for several reasons. In order to construct \emph{kinematic boundary conditions} for the present stress-gradient homogenization, it has firstly be noted that the kinematic micro-macro relations~\eqref{eq:micromacrostrain} and \eqref{eq:micromacromicrodispl} can be transformed to pure surface integrals. 
Thus, it is possible at all to prescribe $\strainmacroSGcomp_{ij}$ and $\microdispldevcomp_{ijk}$ exclusively by suitable boundary conditions (in contrast to micromorphic theory, cf.\ e.~g.\ \citep{Huetter2017,Forest1998,Jaenicke2012}). In particular, an additional quadratic term is added to conventional kinematic boundary conditions
\begin{equation}
	 \displcomp_i=\displmacrocomp_i+\strainmacrocomp_{ij}\locmicrorelcomp_j+C_{ijk}\locmicrorelcomp_j\locmicrorelcomp_k
	 \label{eq:kinamticBCansatz}
\end{equation} 
as proposed in \citep{Gologanu1997,Kouznetsova2002}. It can be verified easily that ansatz~\eqref{eq:kinamticBCansatz} satisfies the classical micro-macro relation~\eqref{eq:micromacrostrain} ad hoc. Furthermore, Eq.~\eqref{eq:micromacrogendispl} yields a set of 18 equations for the micro-displacements $\gendisplcomp_{ijk}$ in terms of the 18 independent componentes of $C_{ijk}$. These equations involve the second geometric moment $\geommomcomp_{ij}=\averop{\locmicrorelcomp_i\locmicrorelcomp_j}$.  For simply shaped volume elements, the second geometric moment is a spherical tensor $\geommomcomp_{ij}=\geommomcomp\Kron{
ij}$. In this case, the system of equations for $C_{ijk}$ can be solved, cf.~\citep{Huetter2017}. After reinserting Eq.~\eqref{eq:micromacrogendispl}, the kinematic boundary condition for the stress-gradient theory reads   
\begin{equation}
	 \displcomp_i=\displmacrocomp_i+\strainmacrocomp_{ij}\locmicrorelcomp_j+\frac{1}{2\geommomcomp}\left(\microdispldevcomp_{ijk}+\microdispldevcomp_{ikj}-\microdispldevcomp_{kji}+\frac{1}{\ndim+2}\microdispldevcomp_{mmi}\Kron{jk}\right)\locmicrorelcomp_j\locmicrorelcomp_k\,.
	 \label{eq:kinematicBC}
\end{equation} 
This boundary condition can be inserted to the left-hand side of the generalized Hill-Mandel condition~\eqref{eq:HillMandel}. 
A comparison with the right-hand side of Eq.~\eqref{eq:HillMandel} shows that the kinetic micro-macro relations read
\begin{align}
	\stressmacrocomp_{ij}=&\frac{1}{\domaincell}\ointcell n_k\stresscomp_{k(i}\locmicrorelcomp_{j)} \dsurf=\averop{\stresscomp_{ij}} \label{eq:micromacrostress} \\
	\stressgradcomp_{ijk}=&\frac{1}{2\domaincell\geommomcomp}\ointcell 2n_p\stresscomp_{p(i}\locmicrorelcomp_{j)}\locmicrorelcomp_k-n_p\stresscomp_{pk}\locmicrorelcomp_i\locmicrorelcomp_j
	+\frac{1}{\ndim+2}n_p\left[\stresscomp_{pk}\Kron{ij}-2\frac{\ndim+3}{\ndim+1}\stresscomp_{p(i}\Kron{j)k}\right]\locmicrorelcomp_m\locmicrorelcomp_m \dsurf \nonumber\\
	=&\frac{1}{\geommomcomp}\averop{\stresscomp_{ij}\locmicrorelcomp_k+\frac{1}{\ndim+2}\left(\Kron{ij}\stresscomp_{km}-2\frac{\ndim+3}{\ndim+1}\stresscomp_{(im}\Kron{j)k}\right)\locmicrorelcomp_m} \,.
	\label{eq:micromacrostressgrad}
\end{align}
It can be verified, that the extended static boundary condition~\eqref{eq:staticBCstressgrad} satisfies these kinetic micro-macro relations.
The quadratic deformation modes are illustrated in \figurename{~\ref{fig:cell_microdispl_kinematicBC}} for certain components of the micro-displacement tensor $\microdispldevcomp_{ijk}$.
It seems to be plausible, that the loading and deformation modes in figures~\ref{fig:cell_Rload} and \ref{fig:cell_microdispl_kinematicBC}, respectively, belong to each other.

\emph{Periodic boundary conditions} can be constructed by amending a fluctuation $\fluctuation{\displcomp_i}(\locmicrorelcomp_k)$ to the kinematic boundary condition~\eqref{eq:kinematicBC}
\begin{equation}
	 \displcomp_i=\displmacrocomp_i+ \strainmacrocomp_{ij}\locmicrorelcomp_j+\frac{1}{2\geommomcomp}\left(\microdispldevcomp_{ijk}+\microdispldevcomp_{ikj}-\microdispldevcomp_{kji}+\frac{1}{\ndim+2}\microdispldevcomp_{mmi}\Kron{jk}\right)\locmicrorelcomp_j\locmicrorelcomp_k+\fluctuation{\displcomp_i}(\locmicrorelcomp_k)\,.
	 \label{eq:periodicBC}
\end{equation} 
This fluctuation field is assumed to be periodic
\begin{equation}
  \fluctuation{\displcomp_i}(\locmicrorelcomp_k^+)=\fluctuation{\displcomp_i}(\locmicrorelcomp_k^-)\,.
	\label{eq:periodicfluctuations}
\end{equation}
Therein, $\locmicrorelcomp_k^+$ and $\locmicrorelcomp_k^-$ refer to homologeous points of the boundary $\domaincellbound$, i.~e., to points with opposing normal $n_i(\locmicrorelcomp_k^-)=-n_i(\locmicrorelcomp_k^+)$ as sketched in \figurename{~\ref{fig:periodic_BCs}}. 
In order to formulate a boundary-value problem for the microscopic displacement field $\displcomp_i(\locmicrorelcomp_k)$, the fluctuations are eliminated in Eq.~\eqref{eq:periodicfluctuations} by Eq.~\eqref{eq:periodicBC}, yielding
\begin{equation}
  \displcomp_i(\locmicrorelcomp_k^+)-\displcomp_i(\locmicrorelcomp_k^-)=\strainmacrocomp_{ij}\left(\locmicrorelcomp_j^{+}-\locmicrorelcomp_j^{-}\right)+\frac{1}{2\geommomcomp}\left(\microdispldevcomp_{ijk}+\microdispldevcomp_{ikj}-\microdispldevcomp_{kji}+\frac{1}{\ndim+2}\microdispldevcomp_{mmi}\Kron{jk}\right)\left(\locmicrorelcomp_j^{+}\locmicrorelcomp_k^{+}-\locmicrorelcomp_j^{-}\locmicrorelcomp_k^{-}\right)
	\label{eq:periodicfluctuationseliminated}
\end{equation}
The periodicity of the fluctuation field, Eq.~\eqref{eq:periodicfluctuations} or \eqref{eq:periodicfluctuationseliminated}, satisfies ad hoc the kinematic micro-macro relation~\eqref{eq:micromacrostrain} for the strain, but not Eq.~\eqref{eq:micromacromicrodispl} for the micro-displacements. Thus, Eqs.~\eqref{eq:micromacromicrodispl} and \eqref{eq:periodicfluctuationseliminated} have to be imposed as global constraints at the micro-scale \citep{Huetter2018Cosserat}. For a hyper-elastic material with strain-energy density $\strainenergy(\straincomp_{ij})$, 
the corresponding Lagrangian thus reads
\begin{equation}
		\begin{split}
		\Lagrangefunc=&\averop{\strainenergy}		
		\!-\!\frac{1}{\domaincell[]}\!\!\!\!\!\int\limits_{\domaincellboundhalf[]}\!\!\!\!\!\Lagrangemult[]_i(\locmicrorelcomp^+_p)\left[
	\displcomp_i(\locmicrorelcomp_k^+)\!-\!\displcomp_i(\locmicrorelcomp_k^-)\!-\!\strainmacrocomp_{ij}\!\left(\locmicrorelcomp_j^{+}\!\!-\!\locmicrorelcomp_j^{-}\right)\!-\!\frac{1}{2\geommomcomp}\!\left(\!2\microdispldevcomp_{i(jk)}\!-\microdispldevcomp_{kji}\!+\!\frac{1}{\ndim\!+\!2}\microdispldevcomp_{mmi}\Kron{jk}\!\right)\left(\locmicrorelcomp_j^{+}\locmicrorelcomp_k^{+}\!\!-\locmicrorelcomp_j^{-}\locmicrorelcomp_k^{-}\right)
		\right]\dsurf \\		
		&+\Lagrangemult_{ijk}\!\left[
\microdispldevcomp_{ijk}-\frac{1}{\domaincell}\!\!\!\ointcell \displcomp_{(i} n_{j)}\locmicrorelcomp_k-\frac{1}{\ndim\!+\!1}	\displcomp_{(i} n_{m)}\locmicrorelcomp_m\Kron{jk}-\frac{1}{\ndim\!+\!1}	\displcomp_{(j} n_{m)}\locmicrorelcomp_m\Kron{ik}\dsurf
			\right]
\,.
		\end{split}
		\label{eq:LagrangefuncminconstrperiodicBC}
\end{equation}
Therein, the first surface integral is taken over one half of the boundary $\locmicrorelcomp_k^+\in\domaincellboundhalf[]$ and the respective homologous points $\locmicrorelcomp_k^-$ have to be given as a function in terms of $\locmicrorelcomp_k^+\in\domaincellboundhalf[]$. Correspondingly, the field of scalar Lagrange multipliers $\Lagrangemult[]_i$ is defined in terms of $\locmicrorelcomp^+_p$. The functional $\Lagrangefunc$ is to be optimized with respect to the microscopic displacement field $\displcomp_i(\locmicrorelcomp_k)$ and to the Lagrange multipliers $\Lagrangemult[]_i(\locmicrorelcomp^+_p)$ and $\Lagrangemult_{ijk}$. The corresponding stationarity conditions are the local equilibrium conditions $\stresscomp_{ij,i}=0$ and $\stresscomp_{ij}=\stresscomp_{ji}$, as well as the enforced relations~\eqref{eq:micromacromicrodispl} and \eqref{eq:periodicfluctuationseliminated} and the boundary conditions
\begin{align}
	n_i\sigma_{ij}=\pm {\Lagrangemult[]}_j(\locmicrorelcomp_k) +n_i {\Lagrangemult[]}_{ijk} \locmicrorelcomp_k \,.
	\label{eq:natBCminconstrperiodic}
\end{align}
The plus sign $+{\Lagrangemult[]}_j(\locmicrorelcomp_k)$ in the first term applies to points $\locmicrorelcomp_k\in\domaincellboundhalf[]$, whereas the minus sign applies to respective homologeous points $\locmicrorelcomp^-_k$.
Thus, the tractions at the boundary, Eq.~\eqref{eq:natBCminconstrperiodic}, involve the anti-periodic part ${\Lagrangemult[]}_j(\locmicrorelcomp_k)$ with an superimposed linear term with ${\Lagrangemult[]}_{ijk}$. 
Correspondingly, the classic case is recovered in absence of stress-gradients. 
For irreversible material behavior, the stationarity conditions are generalized to hold without existence of a Lagrangian function $\Lagrangefunc$ (principle of virtual power).

Insering Eq.~\eqref{eq:natBCminconstrperiodic} to the kinetic micro-macro relations~\eqref{eq:micromacrostress} and \eqref{eq:micromacrostressgrad} yields
\begin{align}
\stressmacrocomp_{ij}=&%
\frac{1}{\domaincell[]}\!\!\!\!\int\limits_{\domaincellbound[]^+}\!\!\!\! \left(\locmicrorelcomp^+_{(i} -\locmicrorelcomp^-_{(i}\right) {\Lagrangemult[]}_{j)}(\locmicrorelcomp^+_k)\dsurf \\
\stressgradcomp_{ijk}=&	{\Lagrangemult[]}_{ijk}
                        \!+\!	\frac{1}{2\domaincell\geommomcomp}\!\!\!\!\int\limits_{\domaincellbound[]^+}\!\!\!\!\! 2 {\Lagrangemult[]}_{(i}\left(\locmicrorelcomp^+_{j)}\locmicrorelcomp^+_k\!\!-\locmicrorelcomp^-_{j)}\locmicrorelcomp^-_k\right)\!-\!{\Lagrangemult[]}_{k}\left(\locmicrorelcomp^+_i\locmicrorelcomp^+_j\!\!-\locmicrorelcomp^-_i\locmicrorelcomp^-_j\right)\!+\!
												\frac{1}{\ndim\!+\!2}\!\left(\!{\Lagrangemult[]}_{k}\Kron{ij}\!-\!2\frac{\ndim\!+\!3}{\ndim\!+\!1}{\Lagrangemult[]}_{(i}\Kron{j)k}\!\right)\left(\locmicrorelcomp^+_m\locmicrorelcomp^+_m\!\!-\locmicrorelcomp^-_m\locmicrorelcomp^-_m\right)
												\dsurf
												\label{eq:stressgrad_periodicBC}
\end{align}
These terms coincide with the coefficients of $\strainmacroSGratecomp_{ij}$ and $\microdispldevratecomp_{ijk}$ when evaluating the left-hand side of the generalized Hill-Mandel condition~\eqref{eq:HillMandel}, so that the latter is satisfied. 

\section{Homogenization of an elastic porous medium}
\label{sec:homogenization_foam}

\begin{figure}
	\centering
	\begin{minipage}{0.46\textwidth}
		\inputsvg{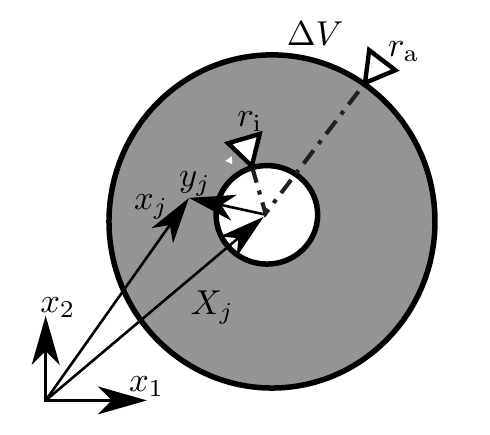}
		\caption{Circular volume element with pore}
		\label{fig:circularvolumeelement}
	\end{minipage}
	\hfill
	\begin{minipage}{0.46\textwidth}
			\inputsvg{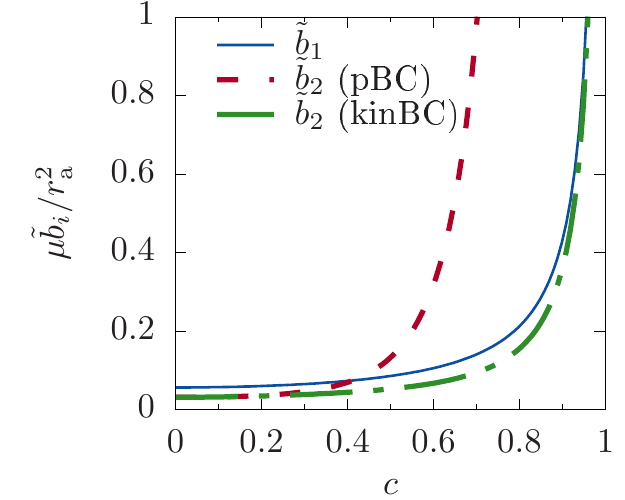}
		\caption{Dependence of stress-gradient compliance moduli on porosity and type (periodic \enquote{pBC} or kinematic \enquote{kinBC}) of boundary conditions ($\Poissrat=0.3$)}
		\label{fig:SGcoeffcompl}
	\end{minipage}
\end{figure}

A circular (or spherical) volume element as shown in \figurename{~\ref{fig:circularvolumeelement}} can be used as approximation to a material with a regular hexagonal arrangement of pores.
The circular volume element has firstly the advantage, that this geometry does not posses preferred directions. Consequently, isotropic behavior of the microscopic constituents will result in an isotropic homogenized behavior. Secondly, certain analytical solutions can be found for this simple geometry. That is why, this geometry has been used within numerous studies on fundamental aspects of homogenization, e.~g.\ \citep{Gologanu1997,Huetter2018Cosserat,Muehlich2012}.

In the present study, the effective properties of the stress-gradient continuum shall be computed for linear elastic behavior $\stresscomp_{ij}=\Lamefirst \Kron{ij} \straincomp_{kk}+2\Lamesec\straincomp_{ij}$ of the matrix material $\radvoid\leq |\locmicrorelcomp_i|\leq\radcell$ using periodic boundary conditions. For the circular volume element, the homologeous points are located opposite to each other $\locmicrorelcomp^-_i=-\locmicrorelcomp^+_i$ with respect to the center of the volume element. Thus, Eq.~\eqref{eq:stressgrad_periodicBC} reduces to $\stressgradcomp_{ijk}={\Lagrangemult[]}_{ijk}$. 
Effectively, this means that the problem~\eqref{eq:LagrangefuncminconstrperiodicBC} can be interpreted as superposition of static boundary conditions for the stress-gradient terms with the conventional periodic conditions for classical behavior, i.~e., for the effective Lamé's constants $\Lamefirsteff$ and $\shearmodeff$ in a relation 
\begin{equation}
	\stressmacrocomp_{ij}=\Lamefirsteff\Kron{ij}\strainmacroSGcomp_{kk}+2\shearmodeff\strainmacroSGcomp_{ij}\,.
	\label{eq:stressstrainSG}
\end{equation}
The solution for $\Lamefirsteff$ and $\shearmodeff$ is well-known. It remains to address the non-classical terms.

In the plane case, the stress gradient tensor has four independent components $\stressgradcomp_{111}=-\stressgradcomp_{122}=-\stressgradcomp_{212}$, $\stressgradcomp_{221}$, $\stressgradcomp_{222}=-\stressgradcomp_{211}=-\stressgradcomp_{121}$, $\stressgradcomp_{112}$, and so does have the tensor of micro-displacements $\microdispldevcomp_{ijk}$ \citep{Forest2012}. 

Favorably, the circular volume element is treated in polar coordinates $r$, $\varphi$. 
In particular, the part of the boundary condition~\eqref{eq:natBCminconstrperiodic}, which is related to the stress gradient, reads
\begin{align}
\stresscomp_{rr}(\radcell)=&\frac{\radcell}{4}\left[-(\stressgradcomp_{221}\!-\!3\stressgradcomp_{111})\cos(3\varphi)
\!+\!(\stressgradcomp_{111}\!+\!\stressgradcomp_{221})\cos(\varphi)
\!+\!(\stressgradcomp_{112}\!+\!\stressgradcomp_{222})\sin(\varphi)
\!+\!(\stressgradcomp_{112}\!-\!3\stressgradcomp_{222})\sin(3\varphi)\right]
\label{eq:BCpolarrr} \\
\stresscomp_{r\varphi}(\radcell)=&\frac{\radcell}{4}\left[(\stressgradcomp_{221}\!-\!3\stressgradcomp_{111})\sin(3\varphi)
\!+\!(\stressgradcomp_{111}\!+\!\stressgradcomp_{221})\sin(\varphi)
\!-\!(\stressgradcomp_{112}\!+\!\stressgradcomp_{222})\cos(\varphi)
\!+\!(\stressgradcomp_{112}\!-\!3\stressgradcomp_{222} )\cos(3\varphi)\right]  
\label{eq:BCpolarrphi}
\end{align}
The problem can be solved with an ansatz for the Airy stress function $\Airyfunc(r,\varphi)$, which involves respective terms of the Mitchell series:
\begin{equation}
\begin{split}
\Airyfunc=&\left[(\stressgradcomp_{111}+\stressgradcomp_{221})\cos(\varphi)+(\stressgradcomp_{112}+\stressgradcomp_{222})\sin(\varphi)\right]\left(A_1 r^3+\frac{A_2}{r}\right) \\
  &+\left[(\stressgradcomp_{221}-3\stressgradcomp_{111})\cos(3\varphi)-\sin(3\varphi)(\stressgradcomp_{112}-3\stressgradcomp_{222})\right]\left(A_3 r^5+\frac{A_4}{r}+A_5 r^3+\frac{A_6}{r^{3}}\right)\,.
\end{split}
\label{eq:Airyansatz}
\end{equation}
The coefficients $A_1$ to $A_6$ can be determined from boundary conditions~\eqref{eq:BCpolarrr} and \eqref{eq:BCpolarrphi}, and the trivial natural boundary condition $\stresscomp_{rr}(\radvoid)=\stresscomp_{r\varphi}(\radvoid)=0$ at the surface of the pore. Instead of evaluating the kinematic micro-macro relation~\eqref{eq:micromacromicrodispl}, the corresponding micro-displacements can be computed equivalently by Castigliano's method. For this purpose, the complementary strain energy is computed as
\begin{equation}
	\strainenergycompmacro\!
	=\averop{\!\frac{1}{4\Lamesec}\!\left(\stresscomp_{ij}\stresscomp_{ij}\!-\!\Poissrat\stresscomp_{kk}^2\right)\!}
	=\frac{\complcoeffSG_1}{2}\!\left[(\stressgradcomp_{111}\!+\!\stressgradcomp_{221})^2\!+\!(\stressgradcomp_{112}\!+\!\stressgradcomp_{222})^2\right]
+\frac{\complcoeffSG_2}{2}\!\left[(\stressgradcomp_{221}\!-\!3\stressgradcomp_{111})^2\!+\!(\stressgradcomp_{112}\!-\!3\stressgradcomp_{222})^2\right]
\end{equation}
with
\begin{align}
	\complcoeffSG_1&=\frac{\radcell^2}{32\Lamesec}\,\frac{3-4\Poissrat+\VVF^2}{1-\VVF^2}\,, &
	\complcoeffSG_2&=\frac{\radcell^2}{32\Lamesec}\,
\frac{1+\VVF+9\VVF^3-7\VVF^2+(3-4\Poissrat)(1+\VVF)(1+\VVF^2)\VVF^2}{(1+4\VVF+\VVF^2)(1-\VVF)^3}
	\label{eq:SGcomplmoduli}
\end{align}
for the plane strain case. Therein, $\VVF=\radvoid^2/\radcell^2$ refers to the porosity of the material. The procedure can be performed analogously for kinematic boundary conditions as described in the appendix. The resulting compliance moduli are plotted in \figurename{~\ref{fig:SGcoeffcompl}}. 
Plausibly, their value tends to infinity as $\VVF$ tends to 1.
The same $\complcoeffSG_1$ is obtained for both types of boundary conditions. The values of $\complcoeffSG_2$ coincide for homogeneous material $\VVF=0$, but for porous material $\VVF>0$ periodic boundary conditions yield more compliant behavior than kinematic boundary conditions. This behavior is known from classical homogenization.
For the plane stress case, $\Poissrat$ in Eq.~\ref{eq:SGcomplmoduli} has to be replaced by $\Poissrat/(1+\Poissrat)$. 

\citet{Forest2012} wrote the non-classical linear-elastic constitutive relation of an isotropic and centro-symmetric material in compliance form in a Voigt-type notation as
\begin{align}
\begin{pmatrix}
	3\microdispldevcomp_{111}\\
	\microdispldevcomp_{221}
\end{pmatrix}
&=\complmatSG\matprod
\begin{pmatrix}
	\stressgradcomp_{111}\\
	\stressgradcomp_{221}
\end{pmatrix}
\,,&
\begin{pmatrix}
	3\microdispldevcomp_{222}\\
	\microdispldevcomp_{112}
\end{pmatrix}
&=\complmatSG\matprod
\begin{pmatrix}
	\stressgradcomp_{222}\\
	\stressgradcomp_{112}
\end{pmatrix}
\label{eq:linearelastcompliance}
\end{align}
The factor 3 in front of $\microdispldevcomp_{111}$ and $\microdispldevcomp_{222}$ was introduced such that the Voigt-type column vectors are work-conjugate to each other. Correspondingly, $\complmatSG$ is a symmetric and positive definite compliance matrix. 
A comparison of Eq.~\ref{eq:SGcomplmoduli} with Eq.~\eqref{eq:linearelastcompliance} shows, that the compliance matrix for the stress gradients has to be identified as
\begin{equation}
	\complmatSG=\begin{pmatrix}
	            	\complcoeffSG_1+9\complcoeffSG_2 &  \complcoeffSG_1-3\complcoeffSG_2 \\
								\complcoeffSG_1-3\complcoeffSG_2 &  \complcoeffSG_1+\complcoeffSG_2
	            \end{pmatrix}\,.
	\label{eq:eq:linearelastcompliancematrixhomogenized}
\end{equation}

\section{Uni-axial tension}
\label{sec:uniaxial_tension}

\subsection{Stress gradient theory}

\begin{figure}
	\centering
	\inputsvg{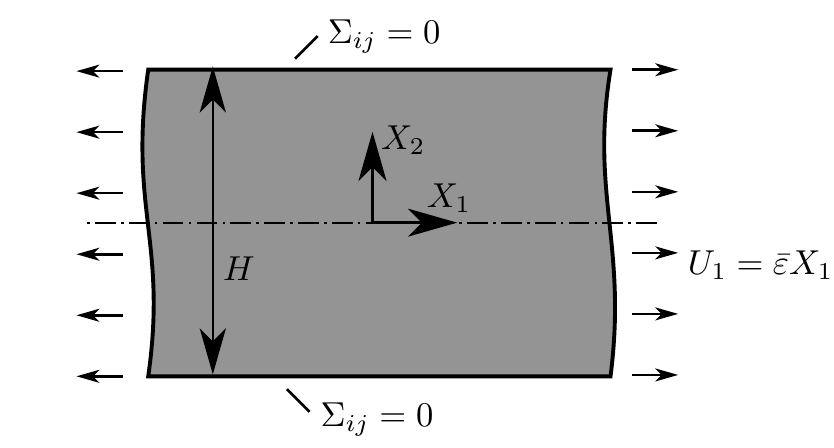}
	\caption{Tensile test with stress-gradient material}
	\label{fig:tensiletest}
\end{figure}

As an example, the predictions of the stress-gradient theory for uni-axial tension shall be investigated as sketched in \figurename~\ref{fig:tensiletest}.
The stress-gradient theory requires extended boundary conditions in form of a second order tensor, cf.~\citep{Forest2012}.
Here, the trivial natural boundary condition 
\begin{equation}
	\stressmacrocomp_{ij}(\locmacrocomp_2=\pm \tenstestpecwidth/2)=0
	\label{eq:BCtensionlatsurface}
\end{equation}
is prescribed at the lateral free surfaces. Consequently, a state of constant stress $\stressmacrocomp_{11}=\mathrm{const.}$ is \emph{not} a solution to the uni-axial tension problem since it would violate the boundary condition~\eqref{eq:BCtensionlatsurface},
in contrast to classical Cauchy continuum theory or even (first order) micro-morphic or strain-gradient theories.

\begin{figure}%
	\centering
	\subfloat[]{\inputgnuplot{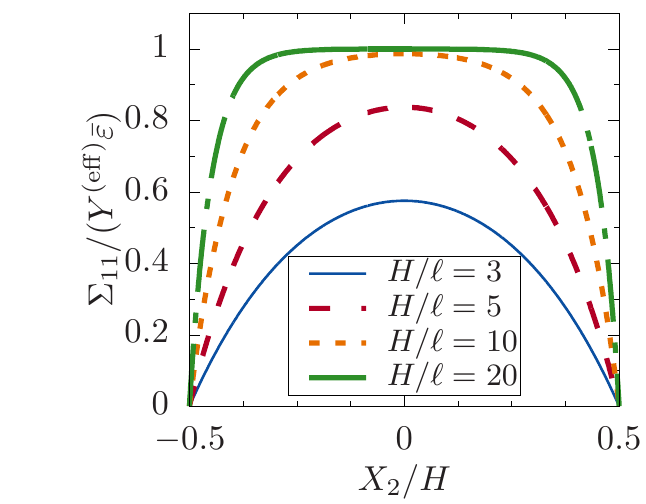} \label{fig:tensteststressdistribution}} 
	\hfill
  \subfloat[]{\inputgnuplot{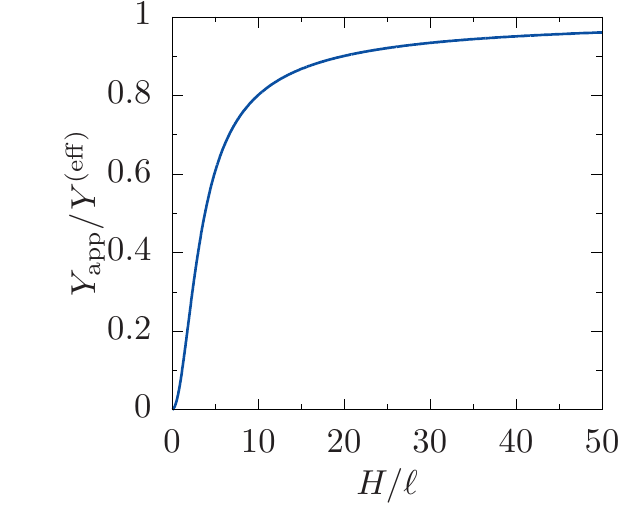} \label{fig:tenstestsizeeffect}}
	\caption[]{Stress-gradient medium under uni-axial tension: \subref{fig:tensteststressdistribution} stresses over cross section, \subref{fig:tenstestsizeeffect} size effect in apparent Young's modulus}%
	\label{fig:tenstestresults}%
\end{figure}

For a sufficiently long specimen, the stress state depends only on $\locmacrocomp_2$ and the only non-vanishing components of stress and its gradient are $\stressmacrocomp_{11}(\locmacrocomp_2)$ and $\stressgradcomp_{112}(\locmacrocomp_2)$, respectively. Inserting the latter to the constitutive relation~\eqref{eq:linearelastcompliance} yields 
$\microdispldevcomp_{222}
=1/3\complmatSGcomp_{12}\stressgradcomp_{112}$ and $\microdispldevcomp_{112}=\complmatSGcomp_{22}\stressgradcomp_{112}$. Correspondingly, the components of the strain tensor, Eq.~\eqref{eq:strainmacro}, are
\begin{align}
	\strainmacroSGcomp_{11}=&\displmacrocomp_{1,1}+\microdispldevcomp_{112,2}\,, &
	\strainmacroSGcomp_{22}=&\displmacrocomp_{2,2}+\microdispldevcomp_{222,2}\,.
	\label{eq:strainmacrotenstest}
\end{align}
Therein , $\displmacrocomp_{1,1}$ equals the applied strain $\straintenstest$. Furthermore, the constitutive relation~\eqref{eq:stressstrainSG} between $\stressmacrocomp_{11}$ and strains $\strainmacroSGcomp_{11}$ and $\strainmacroSGcomp_{22}$ is required. 
Favorably, it is used in compliance form $\strainmacroSGcomp_{11}=\stressmacrocomp_{11}/\emodeff$, wherein $\emodeff$ refers to (macroscopic) Young's modulus. Together with the constitutive law for $\microdispldevcomp_{112}$, Eq.~\eqref{eq:strainmacrotenstest}$_1$ yields the ODE
\begin{equation}
	\stressmacrocomp_{11}-\emodeff\complmatSGcomp_{22}\stressmacrocomp_{11,22}=\emodeff\straintenstest\,,
\label{eq:ODEstresstenstest}
\end{equation}
whose coefficient introduces the intrinsic length $\charlengthSGtens=\sqrt{\emodeff\complmatSGcomp_{22}}$.
Under boundary conditions~\eqref{eq:BCtensionlatsurface}, the solution is
\begin{equation}
	\stressmacrocomp_{11}=\emodeff\straintenstest\left[1-\frac{\cosh\left(\frac{\locmacrocomp_2}{\charlengthSGtens}\right)}{\cosh\left(\frac{\tenstestpecwidth}{2\charlengthSGtens}\right)}\right]
\label{eq:stressmacrotenstest}
\end{equation}
as plotted in \figurename~\ref{fig:tensteststressdistribution} for some parameter sets.
Subsequently, Eq.~\eqref{eq:strainmacrotenstest}$_2$ could be solved for the lateral displacements $\displmacrocomp_{2}(\locmacrocomp_2)$.
Finally, the strain energy within a single cross section $\locmacrocomp_1=\mathrm{const}$ is computed as
\begin{align}
	\frac12\int\limits_{-\tenstestpecwidth/2}^{\tenstestpecwidth/2} \stressmacrocomp_{11}\strainmacroSGcomp_{11}+\stressgradcomp_{112} \microdispldevcomp_{112} \d \locmacrocomp_2 
	&= \frac12 \straintenstest^2	\tenstestpecwidth \underbrace{\emodeff \left[1-\frac{2\charlengthSGtens}{\tenstestpecwidth}\tanh\left(\frac{\tenstestpecwidth}{2\charlengthSGtens}\right)\right]}_{=:\emodapparaent}\,,
	\label{eq:tenstestapparentYoungsmodulus}
\end{align}
from which the apparent Young's modulus $\emodapparaent$ of the specimen can be extracted. The square bracket in Eq.~\eqref{eq:tenstestapparentYoungsmodulus} reflects the size effect. Figure~\ref{fig:tenstestsizeeffect} shows that the apparent Young's modulus of smaller samples is smaller than that of sufficiently large samples. Such negative size effects have already been observed for the stress-gradient continuum under different loading conditions \citep{Tran2018}.

\begin{figure}
	\begin{minipage}[t]{0.46\textwidth}
		\centering
		\inputsvg{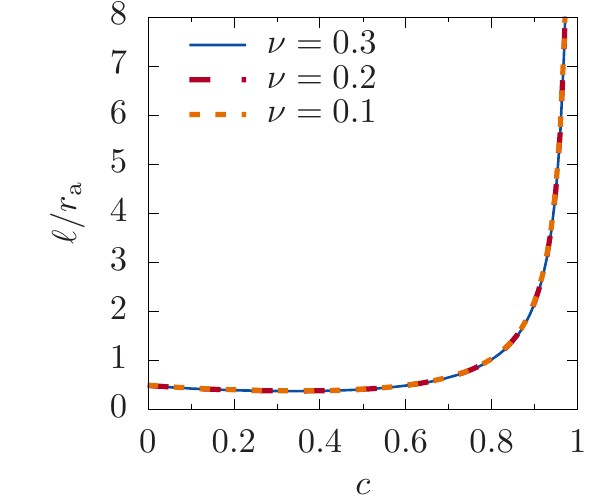}
		\caption{Intrinsic length of stress-gradient theory from homogenization (plane stress)}
		\label{fig:SGintlengthtens}
	\end{minipage} \hfill
	\begin{minipage}[t]{0.46\textwidth}
		\centering
		\inputsvg{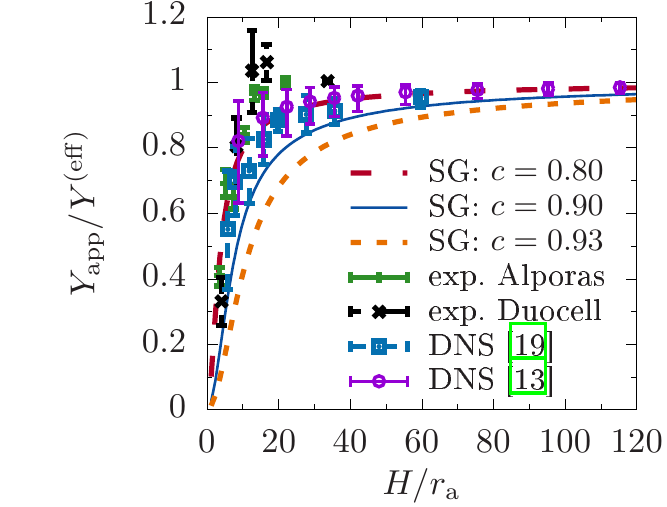}
		\caption{Predictions of stress-gradient theory in comparison with experimental results \citep{Andrews2001} and beam models \citep{Tekoglu2011} of foams ($\Poissrat=0$)}
		\label{fig:cmp_SG_Exp}
	\end{minipage}
\end{figure}

The size effect depends on the single intrinsic length $\charlengthSGtens$ only. The predicted values of this intrinsic length from the homogenization in Section~\ref{sec:homogenization_foam} are depicted in \figurename~\ref{fig:SGintlengthtens}. Thereby, the required effective value of Young's modulus $\emodeff$ from \citep{Huetter2018Cosserat} has been used to compute $\charlengthSGtens$.
The figure shows firstly that Poisson's ratio $\Poissrat$ of the matrix material has a very weak influence on $\charlengthSGtens$.
Secondly, the intrinsic length has an approximately constant and small value $\charlengthSGtens\approx0.5\radcell$ for small porosities $\VVF\lesssim0.6$. 
For larger values of $\VVF$, the value of $\charlengthSGtens$ increases strongly and even tends to infinity as $\VVF$ goes to one. This behavior is attributed to the fact that the classical properties like $\emodeff$ tend to zero as $1-\VVF$, whereas the stress gradient compliance, Eq.~\eqref{eq:SGcomplmoduli}, has a $(1-\VVF)^3$ singularity.
Furthermore, \figurename~\ref{fig:SGintlengthtens} shows that the predicted size effect does not vanish completely for homogeneous material $\VVF=0$. Though, this was neither the case for the strain-gradient theory \citep{Muehlich2012,Gologanu1997,Kouznetsova2004a}. 

\subsection{Comparison with experiments and direct numerical simulations}

It is known that foam materials exhibit size effects when the specimen size becomes comparable to the cell size of the foam. In particular, negative size effects under uni-axial loading have been observed in experiments with foams \citep{Andrews2001} and direct numerical simulations with discretely resolved strut structure \citep{Tekoglu2011, Liebenstein2018}. 
The observed negative size effect was attributed to a surface layer of incomplete cells which do not carry any load \citep{Andrews2001,Wheel2015,Rueger-Lakes2019}.
This surface layer can be seen as physical explanation of the boundary condition~\eqref{eq:BCtensionlatsurface} for the stress-gradient theory in the previous section. It was shown that the stress-gradient theory can describe the negative size effect \emph{qualitatively}. 
The subsequent question is whether the present homogenization approach allows \emph{quantitative} predictions of this size effect. 
Figure~\ref{fig:cmp_SG_Exp} compares the experimental results of \citet{Andrews2001} and the direct numerical simulations (DNS) of \citet{Tekoglu2011} and \citet{Liebenstein2018} with the predictions of the present homogenization theory. \citet{Andrews2001} investigated two materials (\enquote{Alporas}, \enquote{Duocell}). \citet{Tekoglu2011} modeled these foams by plane, Voronoi-tesselated beam networks. They specified a \enquote{cell size $d$}, which is taken here as $d\approx2\radcell$. \citet{Liebenstein2018} investigated honeycomb structures for which $\radcell$ is identified with the radius of a circle of equal area. 
The relative density of the foams was specified to be 7--10\%, corresponding to a porosity of $\VVF=0.90\dots0.93$.
Figure~\ref{fig:cmp_SG_Exp} shows that the trend of the experimental data and direct numerical simulations is captured quite well by the present homogenized stress-gradient theory (\enquote{SG}). 

However, the absolute size effect is slightly overestimated by the homogenized theory if the actual porosity is used. Rather, the average of experimental results and direct numerical simulations comply best with the predictions of the present stress-gradient approach for $\VVF\approx0.80$.
This deviations might be attributed to the simple representation of the pores by circles.
Further studies are required on the effect of the topology of foam micro-structures on size effects during elastic and inelastic deformations.

\section{Summary and conclusions}
\label{sec:summary}

The stress-gradient theory requires a constitutive relation between the tensor of micro-displacements $\microdispldevcomp_{ijk}$ and the stress gradient $\stressgradcomp_{ijk}$. In the present contribution, a homogenization framework was developed to identify this constitutive relation from the microstructure of a material. For this purpose, the static boundary conditions of classical homogenization have been interpreted as a Taylor series, whose subsequent term involves the stress gradient. A condition of macro-homogeneity (generalized Hill-Mandel condition) yields a kinematic micro-macro relation for $\microdispldevcomp_{ijk}$. It turned out that $\microdispldevcomp_{ijk}$ can be identified with the deviatoric part of the first moment of the microscopic strain field. Based on the kinematic micro-macro relations, kinematic boundary conditions for the micro-scale have been identified, where the micro-displacements $\microdispldevcomp_{ijk}$ appear as coefficients of the non-classical quadratic term. 
Furthermore, generalized periodic boundary conditions have been formulated.
The proposed homogenization framework with generalized boundary conditions at the micro-scale and micro-macro relations for all involved kinematic and kinematic quantities allows to use linear or nonlinear constitutive relations at the micro-scale. It is thus well-suited for numerical implementations like FE$^2$.

The homogenization procedure was employed to compute the stress-gradient parameters of an elastic material with pores. These parameters were used to predict the negative size effect of foam materials under uni-axial loading. A comparison with respective experiments and direct numerical simulations from literature exhibited a reasonable agreement.

It shall be mentioned that similar non-classical terms in static or kinematic boundary conditions appear in homogenization approaches towards strain-gradient or (first order) micromorphic theories \citep{Gologanu1997,Huetter2017}. The latter theories predict positive size effects, in contrast to the stress-gradient theory. This means that the choice of the generalized continuum theory to be used at the macro-scale, is an important constitutive assumption itself.   

\section*{Acknowledgments}

The financial support by the Deutsche Forschungsgemeinschaft (DFG) under contract HU~2279/2-1 (GH) is gratefully acknowledged.

\bibliographystyle{elsarticle-harv}
\bibliography{StressGradHomogenisation}

\begin{thebibliography}{21}
\expandafter\ifx\csname natexlab\endcsname\relax\def\natexlab#1{#1}\fi
\expandafter\ifx\csname url\endcsname\relax
  \def\url#1{\texttt{#1}}\fi
\expandafter\ifx\csname urlprefix\endcsname\relax\def\urlprefix{URL }\fi

\bibitem[{Aifantis(2003)}]{Aifantis2003}
Aifantis, E., 2003. Update on a class of gradient theories. Mech. Mater.
  35~(3--6), 259--280.

\bibitem[{Andrews et~al.(2001)Andrews, Gioux, Onck, and Gibson}]{Andrews2001}
Andrews, E.~W., Gioux, G., Onck, P., Gibson, L.~J., 2001. Size effects in
  ductile cellular solids. part ii: experimental results. Int. J. Mech. Sci.
  43~(3), 701--713.

\bibitem[{Eringen and Suhubi(1964)}]{Eringen1964}
Eringen, A.~C., Suhubi, E.~S., 1964. Nonlinear theory of simple micro-elastic
  solids--i. Int. J. Eng. Sci. 2~(2), 189--203.

\bibitem[{Forest and Sab(1998)}]{Forest1998}
Forest, S., Sab, K., 1998. Cosserat overall modeling of heterogeneous
  materials. Mech. Res. Commun. 25~(4), 449--454.

\bibitem[{Forest and Sab(2012)}]{Forest2012}
Forest, S., Sab, K., 2012. Stress gradient continuum theory. Mech. Res. Commun.
  40, 16--25.

\bibitem[{Gologanu et~al.(1997)Gologanu, Leblond, Perrin, and
  Devaux}]{Gologanu1997}
Gologanu, M., Leblond, J.~B., Perrin, G., Devaux, J., 1997. Recent extensions
  of {Gurson's} model for porous ductile metals -- part {II}: A {Gurson}-like
  model including the effect of strong gradients of the macroscopic field. In:
  Suquet, P. (Ed.), Continuum micromechanics. No. 377 in {CISM} Courses And
  Lectures. Springer-Verlag, pp. 97--130.

\bibitem[{Hill(1963)}]{Hill1963}
Hill, R., 1963. Elastic properties of reinforced solids: Some theoretical
  principles. J. Mech. Phys. Solids 11~(5), 357--372.

\bibitem[{Hütter(2017)}]{Huetter2017}
Hütter, G., 2017. Homogenization of a {Cauchy} continuum towards a
  micromorphic continuum. J. Mech. Phys. Solids 99, 394--408.

\bibitem[{Hütter(2019)}]{Huetter2018Cosserat}
Hütter, G., 2019. On the micro-macro relation for the microdeformation for the
  homogenization of heterogeneous materials towards micromorphic and micropolar
  continua. J. Mech. Phys. Solids 127, 62--79.

\bibitem[{Jänicke and Steeb(2012)}]{Jaenicke2012}
Jänicke, R., Steeb, H., 2012. Minimal loading conditions for higher order
  numerical homogenisation schemes. Arch. Appl. Mech. 82~(8), 1075--1088.

\bibitem[{Kouznetsova et~al.(2002)Kouznetsova, Geers, and
  Brekelmans}]{Kouznetsova2002}
Kouznetsova, V., Geers, M. G.~D., Brekelmans, W. A.~M., 2002. Multi-scale
  constitutive modelling of heterogeneous materials with a gradient-enhanced
  computational homogenization scheme. Int. J. Numer. Meth. Engng. 54~(8),
  1235--1260.

\bibitem[{Kouznetsova et~al.(2004)Kouznetsova, Geers, and
  Brekelmans}]{Kouznetsova2004a}
Kouznetsova, V.~G., Geers, M., Brekelmans, W. A.~M., 2004. Size of a
  representative volume element in a second-order computational homogenization
  framework. Int. J. Multiscale Comput. Eng. 2~(4).

\bibitem[{Liebenstein et~al.(2018)Liebenstein, Sandfeld, and
  Zaiser}]{Liebenstein2018}
Liebenstein, S., Sandfeld, S., Zaiser, M., 2018. Size and disorder effects in
  elasticity of cellular structures: From discrete models to continuum
  representations. Int. J. Solids Struct. 146, 97--116.

\bibitem[{Maugin(2011)}]{Maugin2011}
Maugin, G.~A., 2011. A historical perspective of generalized continuum
  mechanics. In: Altenbach, H., Maugin, G.~A., Erofeev, V. (Eds.), Advanced
  Structured Materials. Vol.~7. Springer Berlin Heidelberg, pp. 3--19.

\bibitem[{Mindlin(1964)}]{Mindlin1964}
Mindlin, R.~D., 1964. Micro-structure in linear elasticity. Arch. Ration. Mech.
  An. 16~(1), 51--78.

\bibitem[{Mühlich et~al.(2012)Mühlich, Zybell, and Kuna}]{Muehlich2012}
Mühlich, U., Zybell, L., Kuna, M., 2012. Estimation of material properties for
  linear elastic strain gradient effective media. Eur. J. Mech. A-Solid.
  31~(1), 117--130.

\bibitem[{Rueger and Lakes(2018)}]{Rueger-Lakes2019}
Rueger, Z., Lakes, R.~S., 2018. Experimental study of elastic constants of a
  dense foam with weak {Cosserat} coupling. J. Elasticity., in press.

\bibitem[{Sab et~al.(2016)Sab, Legoll, and Forest}]{Sab2016}
Sab, K., Legoll, F., Forest, S., 2016. Stress gradient elasticity theory:
  Existence and uniqueness of solution. J. Elasticity. 123~(2), 179--201.

\bibitem[{Tekoğlu et~al.(2011)Tekoğlu, Gibson, Pardoen, and
  Onck}]{Tekoglu2011}
Tekoğlu, C., Gibson, L., Pardoen, T., Onck, P., 2011. Size effects in foams:
  Experiments and modeling. Prog. Mater. Sci. 56~(2), 109--138.

\bibitem[{Tran et~al.(2018)Tran, Brisard, Guilleminot, and Sab}]{Tran2018}
Tran, V.~P., Brisard, S., Guilleminot, J., Sab, K., 2018. {Mori-Tanaka}
  estimates of the effective elastic properties of stress-gradient composites.
  Int. J. Solids Struct. 146, 55--68.

\bibitem[{Wheel et~al.(2015)Wheel, Frame, and Riches}]{Wheel2015}
Wheel, M.~A., Frame, J.~C., Riches, P.~E., 2015. Is smaller always stiffer?
  {On} size effects in supposedly generalised continua. Int. J. Solids Struct.
  67-68, 84--92.

\end{thebibliography}

\appendix

\section{Solution for kinematic higher-order boundary conditions}

For the circular volume element, kinematic BCs~\eqref{eq:kinematicBC} read
\begin{align}
\displcomp_{r}(\radcell)=&\frac12(3\microdispldevcomp_{111}\!-\!\microdispldevcomp_{221})\cos(3\varphi)
\!+\!(\microdispldevcomp_{111}\!+\!\microdispldevcomp_{221})\cos(\varphi)
\!+\!(\microdispldevcomp_{112}\!+\!\microdispldevcomp_{222})\sin(\varphi)
\!+\!\frac12(\microdispldevcomp_{112}\!-\!3\microdispldevcomp_{222})\sin(3\varphi)
\label{eq:BCkinempolarrr} \\
\displcomp_{\varphi}(\radcell)=&\!-\!\frac12(3\microdispldevcomp_{111}\!-\!\microdispldevcomp_{221})\sin(3\varphi)
\!+\!2(\microdispldevcomp_{111}\!+\!\microdispldevcomp_{221})\sin(\varphi)
\!-\!2(\microdispldevcomp_{112}\!+\!\microdispldevcomp_{222})\cos(\varphi)
\!+\!\frac12(\microdispldevcomp_{112}\!-\!3\microdispldevcomp_{222} )\cos(3\varphi)  
\label{eq:BCkinempolarrphi}
\end{align}
The problem can be solved with an Airy ansatz analogous to Eq.~\eqref{eq:Airyansatz}
\begin{equation}
\begin{split}
\Airyfunc=&2\Lamesec\left[(\microdispldevcomp_{111}+\microdispldevcomp_{221})\cos(\varphi)+(\microdispldevcomp_{112}+\microdispldevcomp_{222})\sin(\varphi)\right]\left(A_1 r^3+\frac{A_2}{r}\right) \\
  &+2\Lamesec\left[(\microdispldevcomp_{221}-3\microdispldevcomp_{111})\cos(3\varphi)-\sin(3\varphi)(\microdispldevcomp_{112}-3\microdispldevcomp_{222})\right]\left(A_3 r^5+\frac{A_4}{r}+A_5 r^3+\frac{A_6}{r^{3}}\right)\,.
\end{split}
\end{equation}
Finally, a macroscopic strain-energy density of
\begin{equation}
	\strainenergycompmacro\!
	=\frac{\stiffcoeffSG_{1\mathrm{kBC}}}{2}\!\left[(\microdispldevcomp_{111}\!+\!\microdispldevcomp_{221})^2\!+\!(\microdispldevcomp_{112}\!+\!\microdispldevcomp_{222})^2\right]
+\frac{\stiffcoeffSG_{2\mathrm{kBC}}}{2}\!\left[(\microdispldevcomp_{221}\!-\!3\microdispldevcomp_{111})^2\!+\!(\microdispldevcomp_{112}\!-\!3\microdispldevcomp_{222})^2\right]
\end{equation}
is obtained for the plane strain case with
\begin{align}
	\stiffcoeffSG_{1\mathrm{kBC}}&=\frac{18\Lamesec}{\radcell^2}\,\frac{1-\VVF^2}{3-4\Poissrat+\VVF^2}\,, &
	\stiffcoeffSG_{2\mathrm{kBC}}&=\frac{2\Lamesec}{\radcell^2}\,
	\frac{(1-\VVF)\left[(3-4\Poissrat)(1+\VVF)+8\VVF^2-8\VVF^3+\VVF^4+\VVF^5\right]}{(1+\VVF^6)(3-4\Poissrat)+16\VVF^2\Poissrat^2-24\VVF^2\Poissrat+17\VVF^2-16\VVF^3+9\VVF^4}\,.
	\label{eq:SGstiffmodulikinBC}
\end{align}
The respective independent components $\stressgradcomp_{111}$, $\stressgradcomp_{221}$, $\stressgradcomp_{222}$, $\stressgradcomp_{112}$ of the stress gradient are derived by differentiation by respective work conjugate quantities $3\microdispldevcomp_{111}$, $\microdispldevcomp_{221}$, $3\microdispldevcomp_{222}$, $\microdispldevcomp_{112}$, finally yielding a stiffness matrix
\begin{equation}
	\stiffmatSG=\begin{pmatrix}
	            	\frac{\stiffcoeffSG_{1\mathrm{kBC}}}{9}+\stiffcoeffSG_{2\mathrm{kBC}} & \frac{\stiffcoeffSG_{1\mathrm{kBC}}}{3}-\stiffcoeffSG_{2\mathrm{kBC}} \\
								\frac{\stiffcoeffSG_{1\mathrm{kBC}}}{3}-\stiffcoeffSG_{2\mathrm{kBC}} & \stiffcoeffSG_{1\mathrm{kBC}}+\stiffcoeffSG_{2\mathrm{kBC}}
	            \end{pmatrix}\,.
	\label{eq:eq:linearelaststiffnessmatrixhomogenized}
\end{equation}
as inverse $\stiffmatSG=\inv{\complmatSG}$ to the compliance matrix in Eq.~\eqref{eq:linearelastcompliance}.
A comparison with Eq.~\eqref{eq:eq:linearelastcompliancematrixhomogenized} shows that stiffness and compliance coefficients are related by
\begin{align}
	\complcoeffSG_{1}&=\frac{9}{16\stiffcoeffSG_{1}},&
	\complcoeffSG_{2}&=\frac{1}{16\stiffcoeffSG_{2}}\,.
\end{align}

\end{document}